\documentclass[aps,prb,twocolumn,amsmath,amssymb]{revtex4}
\usepackage{times}
\usepackage{graphicx}
\usepackage{amsfonts}
\usepackage{amsmath, amsthm, amssymb}
\usepackage{dsfont}

\newcommand{\xx}{\mathcal{X}} 

\newcommand{\e}[1]{\mathrm{e}^{#1}}

\newcommand{\eg}{\textit{e.g. }}

\def\i{\mathrm{i}}

\begin{document}

\title{Improved Domain-Wall Dynamics and Magnonic Torques 
using Topological Insulators}

\author{Jacob Linder}

\affiliation{Department of Physics, Norwegian University of
Science and Technology, N-7491 Trondheim, Norway}
 
\begin{abstract}
We investigate the magnetization dynamics that arise when a thin-film ferromagnet is deposited on a topological insulator (TI), focusing in particular on domain-wall motion via current and the possibility of a spin-wave torque acting on the magnetization. We show analytically that the coupling between the magnetic domain wall and the TI removes the degeneracy of the wall profile with respect to its chirality and topological charge. Moreover, we find that the threshold for Walker breakdown of domain wall motion is substantially increased and determined by the interaction with the TI, allowing for higher attainable wall velocities than in the conventional case where the hard axis anisotropy determines the Walker threshold. Finally, we show that the allowed modes of spin-wave excitations and the ensuing magnetization dynamics in the presence of a TI coupling enable a magnonic torque acting even on homogeneous magnetization textures. Our results indicate that the TI-ferromagnet interaction has a similar effect on the magnetization dynamics as an intrinsic Dzyaloshinskii-Moriya interaction in ferromagnets.
\end{abstract}

\date{\today}

\maketitle

\textbf{Introduction}. In the mission of finding a viable alternative that is comparable to, and even exceeding, conventional semiconductor/transistor based technology, spintronics has proven itself as a real contender \cite{zutic_rmp_04, brataas_review}. It offers fast read/writing speeds, non-volatility, and remarkably low power consumption when the spin degree of freedom is decoupled from charge. One particularly promising route which may encapsulate all of these qualitites is that of domain wall motion \cite{review_dw}. Transport of spin textures such as domain walls is possible to accomplish both via electric current \cite{slonberger, tatara_prl_04, zhang_prl_04, thiaville_epl_05, yamaguchi_prl_04, vernier_epl_04} and interactions with spin-waves \cite{han_apl_09,yan_prl_11, wang_prl_12}. In order for domain wall motion to serve as a key constituent in spintronics based applications, it is necessary to find ways to obtain rapid transport of such spin textures in a stable way that preserves the topological profile of the domain wall, i.e. without deformation \cite{walker}. In this regard, the prospect of using spin-waves to induce magnetization dynamics has recently garnered a lot of attention due to the reduced dissipation as compared to the Joule heating arising from current-induced domain wall motion.

Driven not only by the prospect of finding ways to optimize magnetization dynamics, but also the considerable interest from a fundamental physics viewpoint, the influence of a topological insulator surface on magnetization dynamics have recently been considered theoretically \cite{burkov_prl_10, garate_prl_10, culcer_prb_10, yokoyama_prb_10, tserkovnyak_prl_12, pesin_natmat_12, cortijo_prb_14}. Most interestingly, a very recent experiment \cite{mellnik_arxiv_14} demonstrated that the spin-transfer torque may be greatly amplified to unprecedented values by using a coupling to a TI. This begs the question if the same could be possible for inhomogeneous magnetic textures such as domain walls. Motivated by this, we consider in this work the coupling between a magnetic material with a domain wall to a topological insulator (TI) and show that this generates several of the desired features mentioned above, thus enhancing the functionality associated with domain wall motion. We demonstrate three key results. First of all, the presence of the TI acts as a stabilizer on the domain wall profile and removes the degeneracy of the domain wall chirality and topological charge. This is similar to what happens in spin-orbit coupled ferromagnets \cite{miron, ryu_prb_12, linder_prb_13, fert_prb_13}, but importantly it \textit{does not require the presence of any electric current} in our case. Secondly, we show that the resulting threshold for Walker breakdown, corresponding to a domain wall which deforms as it propagates, is not only made independent on the hard axis field of the ferromagnet but also quantitatively raised compared to the conventional case by the coupling to the TI. Raising the threshold value is pivotal as it allows for \textit{higher attainable domain wall velocities before deformation sets in}. The fact that it is independent on the hard axis field is of importance since artificially enhancing the Walker threshold by applying a hard-axis field is not a practical solution for devices \cite{miron}. Finally, we demonstrate that the coupling to a TI allows for a spin-wave induced torque even on a homogeneous magnetization texture, in analogy to what is made possible in magnets with a very weak Dzyaloshinskii-Moryia interaction \cite{manchon_arxiv_14}. This remarkable equivalence is a unique feature pertaining to topological insulators and results in the possibility to control the magnetization with magnons in a fashion that \textit{does not require any electric current and thus is accompanied by a very low dissipation}.

\begin{figure}[t!]
\centering
\resizebox{0.5\textwidth}{!}{
\includegraphics{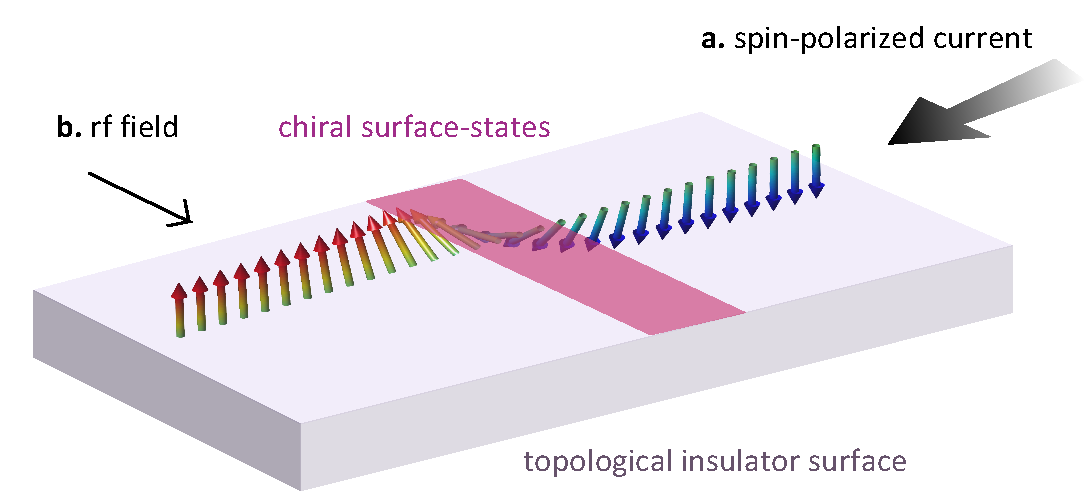}}
\caption{(Color online) Proposed experimental setup: a domain wall ferromagnet is deposited on the surface of a topological insulator and its dynamics can be manipulated via either \textbf{a.} injection of a spin-polarized current or \textbf{b.} propagating spin-waves as excited \eg by an external rf magnetic field. }
\label{model} 
\end{figure}

\textbf{Domain-wall ferromagnet on a topological insulator.} The free energy describing the magnetization texture $\mathbf{m} = \mathbf{M}/M_0$ of a ferromagnet deposited on the surface of a topological insulator (TI) may be written as $F=F_\text{FM} + F_\text{TI}$ where the first term is the free energy of the texture on its own and the second term describes the coupling between the magnetization and the TI. They read $F_\text{FM} = \frac{1}{2} \int \text{d}\mathbf{r} (A\nabla^2\mathbf{m} - K'm_z^2 + K_\perp m_y^2)$, where $A$ is a measure of the exchange stiffness whereas $K'$ and $K_\perp$ denote the easy and hard axis anisotropy energies. Moreover, we have \cite{tserkovnyak_prl_12} $F_\text{TI} = \eta \int \text{d}\mathbf{r} (\mathbf{m}\cdot\nabla m_z) + F'[m_z].$ In the latter expression, $\eta$ describes the coupling between the magnetic texture $\mathbf{m}$ in the FM and the surface-states in the TI whereas the consequence of $F'[m_z]$ is to renormalize the effective anisotropy constant $K' \to K$. One can show that $\eta \propto JJ_\perp$ where $J$ and $J_\perp$ are the in-plane and out-of-plane exchange couplings \cite{tserkovnyak_prl_12}. The equilibrium profile of the domain wall is determined by minimization of the free energy and should also satisfy $\mathbf{m} \times H_\text{eff}=0$ where $H_\text{eff} = - \frac{\delta F}{\delta \mathbf{M}}$ is the effective field and $\delta$ denotes the functional derivative. Note that the out-of-plane domain wall profile is permitted when the combined effort of a strong uniaxial magnetocrystalline anisotropy combined with its effective enhancement due to the coupling term $F'[m_z]$ from the TI together overcome the shape anisotropy field. Let us use denote the tilt angle of the domain wall as $\phi$ and its topological charge as $\sigma$, in which case we find that the equilibrium solution takes the standard form \cite{walker} $\mathbf{m} = (\sin\theta\cos\phi,\sin\theta\cos\phi,\sigma\cos\theta)$, but with one additional criterium: 
\begin{align}
s_\eta\sigma\cos\phi = +1
\end{align}
must be satisfied where $s_\eta = \text{sign}(\eta)$. In the absence of the exchange coupling to the TI, any combination of $\sigma=\pm1$ and $\phi={0,\pi}$ are ground-state solutions, even in the presence of a hard magnetic axis $K_\perp$, but this degeneracy is now lifted such that the TI stabilizes one particular topological profile of the magnetic domain wall.

Such an effect also occurs in recently studied spin-orbit coupled domain wall ferromagnets \cite{miron, ryu_prb_12, linder_prb_13, fert_prb_13}, albeit as a non-equilibrium effect contingent on the presence of an electric current in that case. In contrast, the selection of a particular domain wall profile which is energetically favorable here occurs in equilibrium without any applied spin-polarized current. This has important ramifications for the resulting equations of motion describing domain wall motion, as we now proceed to demonstrate. Assume in what follows for concreteness that $\eta>0$ (our findings are equally valid for $\eta<0$). We employ a collective-coordinate framework in terms of the soft modes $\mathcal{X}$ and $\phi$ of the DW dynamics, where $\mathcal{X}=\mathcal{X}(t)$ is the domain wall center of mass coordinate and $\phi=\phi(t)$ is its tilt angle. As long as $\phi$ remains constant, the domain wall propagates as a rigid soliton and does not deform. Above a certain critical value of the applied current density in the ferromagnet, Walker breakdown sets in as $\dot{\phi}$ becomes non-zero and the domain wall starts to rotate. In order to use domain walls as information carriers in spintronics-based applications, it is desirable to delay the onset of Walker breakdown as much as possible in order to achieve the highest possible domain wall velocities without causing deformation of the magnetization texture. Inserting the above ansatz for $\mathbf{m}$ into the full LLG equation which includes the spin-transfer torque of a current flowing through the ferromagnet, $\partial_t\mathbf{m} = -\gamma\mathbf{m}\times H_\text{eff} + \alpha \mathbf{m} \times \partial_t\mathbf{m} +j\partial_x\mathbf{m} - \beta j\mathbf{m}\times\partial_x\mathbf{m}$ where $j$ is a measure of the applied current density in the ferromagnet and $\theta = 2\arctan[\e{-(x-\mathcal{X})/\Delta}]$ ($\Delta=\sqrt{A/K}$ is the domain wall width), we obtain the following equations of motion after performing an integration over space: 
\begin{align}\label{eq:cc}
\sigma\dot{\xx}+\alpha\dot{\phi} &= -\sigma \tilde{j} - \tilde{K}_\perp\sin2\phi + \sigma\tilde{\eta}\sin\phi,\notag\\
\dot{\phi} - \alpha\sigma \dot{\xx} &= \beta\tilde{j}\sigma.
\end{align}
Here, we have for compactness expressed the equations in terms of normalized coordinates which are all proportional to the original parameters, i.e. $\tilde{j} \propto j, \tilde{K}_\perp \propto K_\perp, \tilde{\eta} \propto \eta$. We will later discuss the explicit magnitude of these terms. Several conclusions may be drawn from the above set of coupled equations. First of all, it is seen that the exchange coupling $\tilde{\eta}$ to the TI induces a term $\propto \sin\phi$ analogously to spin-orbit coupled ferromagnets in the presence of a current \cite{ryu_prb_12, linder_prb_13}. Thus, this term acts as an effective spin-orbit torque, but importantly \textit{it is here present even without application of a current}. This explains why the equilibrium topological profile of the magnetic texture is stabilized only for specific chiralities. To show that this term has important consequences also out-of-equilibrium, i.e. in the presence of an electric current flowing through the ferromagnet, we note that in the region of interest in terms of applications (below Walker threshold, such that $\dot{\phi}=0$) one has: 
\begin{align}\label{eq:walker}
\sigma\alpha\tilde{K}_\perp \sin2\phi = (\beta-\alpha)\tilde{j} + \tilde{\eta}\alpha\sin\phi.
\end{align}
As long as this equation is satisfied, the domain wall will propagate without deformation. Without the exchange coupling $\tilde{\eta}$, it is seen that Walker breakdown will occur when $|\tilde{j}|>|\tilde{j}_c| \equiv |\alpha\tilde{K}_\perp/(\beta-\alpha)|$ \cite{thiaville_epl_05} since the l.h.s. and r.h.s. of the equation no longer has any crossing point when this is satisfied. This changes in the presence of the TI coupling $\tilde{\eta}$. In fact, the criterium which guarantees a solution of Eq. (\ref{eq:walker}) now becomes completely independent on the hard axis field $\tilde{K}_\perp$. When $\alpha\tilde{\eta}$ is larger than $(\beta-\alpha)\tilde{j}$, the l.h.s. and r.h.s. both have maxima and minima with opposite signs and a crossing point is ensured. Thus, there is no Walker breakdown if:
\begin{align}\label{eq:newcond}
|\tilde{j}| < \alpha\tilde{\eta}/|\beta-\alpha|.
\end{align}
We emphasize that this equation is the \textit{lower limit} which guarantees the absence of Walker breakdown, in contrast to the threshold current without the TI coupling which is determined by $\tilde{K}_\perp$ which represents the absolute \textit{upper limit} for the current. In fact, solutions of Eq. (\ref{eq:walker}) could still exist for higher currents, but no analytical expression may be derived for this regime. Eq. (\ref{eq:newcond}) is valid both with and without a hard-axis field. This is advantageous since it relaxes the requirements on the anisotropy properties of ferromagnetic materials used for domain wall motion. As pointed out in Ref. \cite{miron}, artificially enhancing the Walker threshold by applying a hard-axis field is not a practical solution for devices. The coupling to the TI provides a way around this problem. The stronger the exchange coupling $\tilde{\eta}$ between the TI and the FM, the more stable the topological profile of the domain wall becomes and the higher the Walker breakdown limit reaches. The ratio 
\begin{align}\label{eq:R}
\mathcal{R} = \tilde{\eta}/\tilde{K}_\perp = JJ_\perp/(2\pi\hbar v_F\Delta K_\perp d)
\end{align}
is a measure of how effective the TI coupling is with regard to the Walker threshold. Here, $v_F$ is the Fermi velocity of the TI surface and $d$ is the thickness of the ferromagnet. Using $K_\perp = 10^4$ J/m$^3$, $d=2$ nm, $\Delta = 5$ nm, $v_F=10^5$ m/s, we find that $\mathcal{R} \gg 1$ for an exchange coupling $J,J_\perp = 30$ meV (\textit{ab initio} calculations \cite{eremeev_prb_13} of FM/TI interfaces estimate that exchange field gaps as large as $\sim 50$ meV could be obtained for Bi$_2$Se$_3$/MnSe) and that it would continue to rise quadratically with increasing couplings $\{J,J_\perp\}$. The observation of this effect would entail probing the displacement of the domain wall as a function of applied current density as can be experimentally imaged \eg by using a wide-field Kerr microscope \cite{miron}. The occurence of Walker breakdown, with its inherent decrease of domain wall velocity, would then be observed at much higher current densities when the ferromagnetic film is coupled to a TI surface. Ferromagnetic thin films with a thickness down to 1.3 nm displaying a perpendicular anisotropy easy axis have been experimentally realized in Ref. \cite{ikeda_nature_10}. The ratio of the Walker breakdown limits when using proximity-coupling to a TI compared to the conventional hard-axis anisotropy is seen from Eq. (\ref{eq:R}) to scale as $1/K_\perp$, indicating that ferromagnets with a relatively small $K_\perp$ can be made suitable for hosting domain wall motion without suffering from early Walker breakdown by utilizing the TI-coupling. In this way, the effect predicted here could enable use of new types of ferromagnetic materials for domain wall motion which previously were inadequate due to a low Walker threshold.

Let us also comment on the role of including Dzyaloshinskii-Moriya (DM) interactions \cite{DM} that may be generated due to inversion symmetry breaking in thin-film structures. The presence of DM-interactions could be expected to act as a chirality-selector similarly to the TI exchange coupling and thus enhance the predicted effects. This is supported by the results reported in Ref. \cite{thiaville_epl_12} which studied the role of DM-interactions on domain wall motion in a thin-film setup, albeit without any coupling to a TI, where it was found under simplifying assumptions that an extra term $\propto D\sin\phi$ appears in the equation for the Walker threshold ($D$ is the DM-interaction magnitude). This would be equivalent to the $\tilde{\eta}\sin\phi$ term derived in Eq. (\ref{eq:walker}) which suggests that DM-interactions would further enhance the Walker threshold limit. It should be noted that for very strong DM-interactions, the domain wall configuration itself becomes energetically disfavorable \cite{tretiakov_prl_10} which is not the case we are interested in for the present manuscript. 

Domain wall motion is also known to be possible via magnons, in effect by inducing spin-waves that propagate through and interact with the domain wall. This can be accomplished both via an external rf magnetic field or via a thermal gradient and renders possible domain wall motion even in electrically insulating systems. A spin-wave perturbation on top of the magnetic domain wall background may be represented via $\mathbf{m} = \mathbf{m}_0 + \e{-\i \omega t}(s_\theta\hat{\theta} + s_\phi\hat{\phi})$ where $\omega$ is the frequency and $\{s_\theta,s_\phi\}$ are the components of the spin-wave transverse to the local magnetization texture defined by $\hat{r} \parallel \mathbf{m}_0$. Inserting this ansatz into the LLG equation, one obtains the following equations of motion for the spin-waves:
\begin{align}\label{eq:spinwaveDW}
\i\omega s_\theta/\gamma &= -A\partial_x^2s_\phi + s_\phi\Delta[A\Delta(2\cos^2\theta_0-1)-\eta\sin\theta_0],\notag\\
\i\omega s_\phi/\gamma &= A\partial_x^2s_\theta - s_\theta A\Delta^2(2\cos^2\theta_0-1).
\end{align}
In the limit $\eta\to 0$, this is consistent with Ref. \cite{yan_prl_11} and the two equations can be combined into a single equation for a wave $s_\theta-\i s_\phi$ satisfying a Schr{\"o}dinger-like equation. In the present case, the exchange coupling to the TI breaks the symmetry between the equations and influences the resulting domain wall motion induced by the magnons.  For an inhomogeneous magnetic texture, the equations above do not offer any transparent analytical solution, but it is clear from Eq. (\ref{eq:spinwaveDW}) that the spin-wave modes $s_{\phi,\theta}$ couple in a very different manner compared to isolated ferromagnets \cite{yan_prl_11}. However, as one moves away from the domain wall center (i.e. for an effective homogeneous ferromagnet coupled to a TI) we show that, remarkably, propagating spin-waves will still act with a spin-transfer torque on the magnetization solely due to the exchange coupling $\eta$. Such an effect is not present for an isolated ferromagnet where $\eta=0$ and thus offers a new way to induce magnetization dynamics via spin-waves. A similar effect was very recently shown to take place in ferromagnets hosting Dzyaloshinskii-Moryia interactions \cite{manchon_arxiv_14}. In our case, no such interactions are required and we demonstrate that the magnon-induced torque is controlled by both the topological charge and the sign of the exchange coupling $\eta$ to the TI.

\textbf{Magnon-induced torque due to the FM/TI coupling.} We set the equilibrium orientation to $\mathbf{m}_0 \parallel \mathbf{z}$ so that the magnetization may be written generally as $\mathbf{m} = (\delta m_x+s_x,\delta m_y+s_y, \sigma)$ with $\sigma=\pm 1$. In comparison with the previous considered case, this corresponds to the magnetization texture away from the domain wall center. Here, $\delta m_{x,y}$ are the changes in the magnetization texture due to the spin-wave perturbations whereas $s_{x,y}$ describe the spin-waves themselves. The spin-waves vary on a much shorter time-scale than the magnetization texture which is slow in comparison \cite{manchon_arxiv_14}. The first step is to obtain an expression for the spin-waves. By inserting $\mathbf{m}$ into the LLG equation, writing the effective field as $\mathbf{H}_\text{eff} = A\partial_x^2\mathbf{m} + Km_z\hat{\mathbf{z}} + \eta(\partial_xm_x\hat{\mathbf{z}} - \partial_xm_z\hat{\mathbf{x}})$, we obtain $\partial_t s_x = -\alpha\sigma\partial_t s_y - \gamma \sigma\epsilon s_y H_k +\gamma\sigma A\partial_x^2 s_y$ and $\partial_t s_y = \alpha\sigma\partial_t s_x + \gamma \sigma s_x H_k - \gamma\sigma A\partial_x^2s_x$ by linearizing the equations in $s_{x,y}$ and dropping higher-order terms in $\delta m_{x,y}$ and $s_{x,y}$.  Combining these into a single equation for a spin-wave $s_\phi \equiv s_x+\i s_y$, we obtain:
\begin{align}
(1-\i\alpha\sigma)\partial_ts_\phi(x,t) = \i H_k\sigma s_\phi(x,t) - \i\gamma\sigma A\partial_x^2s_\phi(x,t).
\end{align}
This is a separable partial differential equation with exact solution $s_\phi(x,t) = (A_0 \e{\sqrt{c}x} + B_0\e{-\sqrt{c}x})c_0\e{kt}$ where $\{A_0,B_0,C_0\}$ are determined from initial conditions and we defined $k = \gamma\sigma(A c - H_k)/(\i+\alpha\sigma)$. The complex constant $c$ may be determined by demanding that this solution describes oscillating spin-waves such that $k=\i\omega$ where $\omega$ is the spin-wave frequency. This means that Re$\{k\}=0$ while Im$\{k\}=\omega$, which gives the solution:
\begin{align}
c = -q^2 + \i\omega\alpha/(\gamma A)
\end{align}
where $q$ is the wavevector of the magnons satisfying $\omega^2 = \gamma^2(Aq^2 + H_k)^2$. The imaginary part of $c$ is related to the decay length of the spin-waves as they propagate away from their source. This can be seen from the spatially dependent part of the spin-wave wavefunction $s_\phi(x,t)$ since when $\alpha\ll1$ we have $\e{\pm\sqrt{c}x} \simeq \e{\pm\i qx \pm x/(2\xi)}$ where $\xi = q\gamma A/(\omega \alpha)$ is the decay length. Note that the sign of the last term in the exponent has to be chosen so that the spin-waves are attenuated away from the source. With the solution for the spin-waves in hand, we can now consider the equations of motion for the magnon-torque induced changes in magnetization $\delta m_{x,y}$. This can be done by averaging the LLG equations over a spin-wave precession period, so that any linear terms in $s_{x,y}$ or their derivatives equal zero. The $\delta m_{x,y}$ are constants over this short time-interval, and one finally arrives at:
\begin{align}\label{eq:m}
A\partial_x^2\delta m_y = H_k\delta m_y + \sigma\eta \langle s_y\partial_x s_x\rangle,\notag\\
A\partial_x^2\delta m_x = H_k\delta m_x + \sigma\eta \langle s_x\partial_x s_x\rangle.
\end{align} 
The averages $\langle \ldots \rangle$ are both proportional to $\e{-|x-x_0|/\xi}$, where $x_0$ is the source where the spin-waves are generated. Eqs. (\ref{eq:m}) have the same functional form as the magnon-induced torques in Ref. \cite{manchon_arxiv_14} and thus generate both a field-like and damping-like torque on the magnetization due to the coupling $\eta$ to the TI. It is remarkable that the exchange interaction between a TI and ferromagnet generates the same type of magnon-induced torque as the Dzyaloshinskii-Moryia interaction. This result is a unique property of how the TI surface couples to the local magnetic moments in the ferromagnetic region due to the spin-momentum locking of the TI surface electrons and does not appear for exchange couplings to conventional metallic systems. The torque increases in magnitude with $\eta$ and vanishes all-together when $\eta\to0$ since the boundary conditions then only allow the trivial solutions $\delta m_{x,y}=0$. The experimental observation of this effect would be constituted \eg by measuring a reoriented (i.e. tilted) magnetization vector of a homogeneous ferromagnet due to the torque induced by the spin-waves, thus offering a way to control the magnetization direction of the system.

\textbf{Conclusion.} In summary, we have demonstrated that the exchange coupling between a textured ferromagnet and the surface of a topological insulator (TI) provides a venue for improved domain wall motion and spin-transfer torques, both via current-induced motion and spin-waves. In the current-induced case, we found that the coupling to the TI acts as a selector for specific topological profiles of the domain wall and that it rendered the Walker threshold insensitive to the hard-axis anisotropy in the ferromagnet which could be advantageous for practical applications. In particular, it could enable use of new types of ferromagnets to host domain wall motion which previously were inadequate due to a small hard axis anisotropy which in turn would lead to an early onset Walker breakdown. Moreover, we demonstrated that the allowed spin-wave modes propagating through a domain wall would be qualitatively altered compared to an isolated ferromagnet as the coupling to the TI breaks the symmetry that allows formation of circularly polarized waves. On the other hand, spin-waves propagating sufficiently far away from the domain wall center where the magnetization is homogeneous would induce a torque which is proportional to the strength of the TI exchange coupling, thus offering a new way to induce magnetization dynamics via spin-waves by utilizing topological insulators. Our results indicate that the TI-ferromagnet interaction has a similar effect on the magnetization dynamics as an intrinsic Dzyaloshinskii-Moriya interaction in ferromagnets.

\textbf{Acknowledgments}. This work was supported by the Research Council of Norway, Grant No. 205591/F20 (FRINAT), and COST Action MP-1201 "Novel Functionalities through Optimized Confinement of Condensate and Fields".

\end{document}